\documentclass[a4paper,11pt]{article}
\pdfoutput=1 

\usepackage{jheppub} 

\usepackage{graphicx}
\usepackage{amsmath}
\usepackage{amssymb}
\usepackage{amsfonts}
\usepackage{color}
\usepackage{array}

\title{Fast Radio Bursts and    Electromagnetic Transition Radiation  on Gravitational Shockwaves}

\author[a]{D.V. Fursaev,}
\author[a,b]{E.A. Davydov,}
\author[a,b]{and V.A. Tainov}

\affiliation[a]{Bogoliubov Laboratory of Theoretical Physics, Joint Institute for Nuclear Research,\\ 141980, Dubna, Russia}
\affiliation[b]{Dubna State University,\\ 141980, Dubna, Russia,}

\emailAdd{fursaev@theor.jinr.ru}
\emailAdd{davydov@theor.jinr.ru}
\emailAdd{tainov@theor.jinr.ru}

\abstract{When a gravitational shockwave hits a magnetar it creates perturbations of the magnetar magnetic field in a form of a transition radiation.
	We argue that this radiation can be a novel candidate to explain the origin of fast radio bursts (FRB). A unique feature of the transition radiation on the shockwaves is that normal components of its Maxwell strength `remember' only the spatial `profile' of the shock, not the form of the signal. This fact allows us to  determine completely a characteristic initial problem for the perturbations with Cauchy data defined on a null hypersurface just behind the shockwave front. The computations are carried by modeling magnetar as a magnetic dipole. As an illustration we consider shockwaves created by ultrarelativistic objects of two types, by compact sources or by cosmic strings. In the both cases the duration of the engine pulse is determined by an impact distance between the magnetar and the source.  We present the angular distribution of the transition radiation flux and show that it is consistent with properties of the FRB engine.}

\begin{document} 
\maketitle
\flushbottom

\section{Introduction}
\label{sec:intro}

 Fast radio bursts are bright pulses of radio frequencies between 400 MHz and 8 GHz. Although the origins of FRBs are still unknown,
there has been significant progress  in understanding their possible physics~\cite{Petroff:2021wug}  in recent years. Some, and probably all, of these bursts are generated by the release of energy from a magnetar~\cite{Zhang:2022uzl}. This process involves two mechanisms: the primary perturbation of the magnetar's magnetosphere and the subsequent conversion of the perturbation energy into a radio burst. Observational data make it possible to significantly restrict the class of models describing the second mechanism --- the emission of radio waves. One of the most successful models is the synchrotron maser emission~\cite{Lyubarsky:2014jta,Beloborodov:2017juh}. This process occurs at a noticeable distance from the magnetar and is initiated by the primary short intense outburst of energy from the magnetosphere~\cite{Metzger:2019una,Beloborodov:2019wex}. However, the primary perturbation of the magnetosphere can be caused by various reasons. If a blast satisfies some basic criteria~\cite{Margalit:2019tph}, it, regardless of its initial nature, can potentially generate an observed FRB. Therefore, it is important to study all the fundamental mechanisms, the FRB `engines',  that lead to the high-energy outburst from magnetar.

The engines can be divided into two main classes: those caused by internal processes of the magnetar, and those caused by external impact. Among external impacts, mainly collisions are  considered: with asteroids~\cite{Dai:2016qem}, another neutron star~\cite{Moroianu:2022ccs}, black hole~\cite{Kim:2024fuy} or other massive compact objects~\cite{Liu:2020zyj}. Interactions without collision are also considered~\cite{Zhang:2020eou}. The aim of the present work is to demonstrate a novel `collisionless' mechanism of external impact leading to a short powerful outburst of radiation from the magnetar. This is a transition radiation of the magnetar dipole moment under the action of a gravitational shockwave (GSW).

The transition radiation has been described in 1945 by Ginzburg and Frank \cite{Ginzburg:1945zz} as a universal phenomenon
which occurs when a compact charged source moves from one region of space to another with different properties. In the original version, this was the radiation of a uniformly moving point charge crossing the boundary of two media with different permittivities. An alternative scenario is the incidence of a permittivity wave (associated, for example, with a density wave) on a charge at rest~\cite{Ginzburg:1979wi}. Obviously, the solution of Maxwell equations is affected not only by a change in permittivity, but also by a change in the spacetime geometry. Therefore, Ginzburg and Tsytovich subsequently considered the transition radiation that occurs when a weak gravitational wave falls on a compact source of an electromagnetic field~\cite{Ginzburg:1975}. In particular, they studied the scattering of a gravitational wave by a magnetar and the formation of electromagnetic transition radiation as well as Alfven waves in this process. However, for the weak gravitational wave, the resulting radiation also turned out to be too weak to lead to noticeable observable effects. Alternative considerations of energy transfer between gravitational and electromagnetic waves in an external stationary electromagnetic field have repeatedly been encountered in the literature, e.g., the Gertsenshtein effect~\cite{Gertsenshtein:1962},  see also~\cite{Wen:2014wxa}.

A natural way to enhance the power of the transition radiation on a gravitational wave is to consider strong gravitational waves, and
gravitational shockwaves, in particular.
GSW are solutions to the Einstein equations which describe short gravitational pulses with rapid oscillations of the Riemann curvature on the shockwave fronts.  In general, the Einstein and Maxwell equations then become essentially nonlinear in the metric, which leads to significant computational difficulties.  In this work we study plane-fronted shockwaves which propagate in flat space-time or in a space-times where gravitational effects are weak with respect to the gravitational shocks. Such gravitational shockwaves are known to be created by the motion of ultrarelativistic sources and have a unique feature: their Einstein tensor is linear in the metric. These GSW include the subclass of Kundt spacetimes~\cite{Kundt:1961}.

In our approximation, test particles before and after the shockwave pulse move along geodesics in the Minkowsky space-time.
After the shockwave, mutual orientations of 4-velocities of the particles change, as a manifestation of the gravitational memory effect.
Transformations of the 4-velocities on the shockwave front look as coordinate transformations determined by the profile of the GSW.
These transformations are also known as the Penrose supertranslations \cite{Penrose:1972xrn}.
Analogous effects hold for the transition radiation and play the crucial role in the subsequent analysis.

\begin{figure}[t]
	\centering
	\includegraphics[width=0.8\columnwidth]{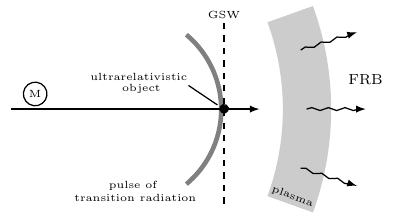}
	\caption{\small  illustrates a novel FRB engine mechanism when an ultrarelativistic object moves near a magnetar $M$. A pulse of a transition radiation is generated by a plane-fronted gravitational shockwave accompanying the source. The pulse then perturbates the plasma surrounding the magnetar, and the pulse energy converts into a fast radio burst. } \label{f1}
\end{figure}

The main results of the present work are two-fold. First, we demonstrate that the gravitational memory effects
on electromagnetic perturbations caused by the considered GSW have a very simple form.
Similar to trajectories of particles the transition radiation `remembers'  the profile of a GSW but does not depend on the complexity of the gravitational `signal'. This fact allows one to overcome computational difficulties relatively easily. Second, on this base, we study perturbations of the magnetar's magnetic field caused by GSW created by different ultrarelativistic objects which move near a magnetar.  We analyze the angular distribution of the transition radiation flux, the duration of the pulses and show that they are consistent with properties of the FRB engines.
Our suggestion for the novel FRB engine mechanism is illustrated  on figure  \ref{f1}.

\section{Gravitational memory of trajectories after shockwaves}   
\label{sec:2}

We start with necessary definitions and illustration of the gravitational memory effect for geodesics on the considered class of plane-fronted shockwave geometries. For a shock moving along
the $x$ direction, from the left to the right, the space-time metric is chosen to be
\begin{equation}\label{1.1}
	ds^2=-dv du -H(v,u,y)du^2+d y_i^2\,.
\end{equation}
The retarded  and advanced time coordinates are  $u=t-x$, $v=t+x$. With vectors $l_\mu=-\delta^u_\mu$, which are null normals to constant $u$ hypersurfaces, the metric tensor can be written as $g_{\mu\nu}=\eta_{\mu\nu}-Hl_\mu l_\nu$, where $\eta_{\mu\nu}$ is the Minkowsky metric.
The properties of the wave are described by $H$ which we take as $H(v,u,y)=\chi(u) f(v,y)$,
where $f(v,y)$ is called the 'profile' function. Evolution of the shock in the retarded time is determined by
$\chi(u)$ which can be called the 'signal' function.  We also define the integrated signal function
\begin{equation}\label{1.3}
	\bar{\chi}(u)=\int^u_{-\infty} \chi(u')du'\,.
\end{equation}
To normalize the profile and signal functions we assume that $\bar{\chi}(u)\to 1$ at $u\to \infty$.

In this paper we consider gravitational perturbations whose signal function is non-vanishing only in some time interval, say, when $0<u<\delta$. Inside the interval $\chi(u)$ can be absolutely arbitrary.
At $u<0$, $u>\delta$ metric \eqref{1.1} coincides with the Minkowsky metric. As has been mentioned above, the important property of
\eqref{1.1}  is that non-vanishing components of its Einstein tensor are linear in $H$,
\begin{equation}\label{1.6}
	G_{uu}=\frac 12 \partial^2_{i} H\,,\quad
	G_{ui}=\partial_v\partial_i H\,,\quad
	G_{ij}= 2\delta_{ij}\partial_v^2 H\,.
\end{equation}
Therefore in the Einstein theory
the shockwaves are produced by sources with the stress-energy tensor
\begin{equation}\label{1.7}
	T_{\mu\nu}=\chi(u)\left (\sigma l_\mu l_\nu+j_i ( l_\mu e_\nu^i+ l_\nu e_\mu^i)+p\delta_{ij}e_\mu^ie_\nu^j\right)\,,
\end{equation}
where 
\begin{equation}
	\sigma=f_{,ii}/\kappa\,,\quad j_i=-2f_{,vi}/\kappa\,, \quad p=4f_{,vv}/\kappa\,,\quad \kappa=16\pi G\,.
\end{equation}
Vectors $e^i_\mu=\delta^i_\mu$ are tangent to null hypersurfaces of constant $u$.  Concrete examples of the profile functions
are given below for ultrarelativistic particles and cosmic strings.

An immediate consequence of \eqref{1.6} is that
$G_{\mu\nu}$ is well-defined in the limit of extremely short signals, when $\chi(u)\to \delta(u)$.  The wave front of GSW is then localized exactly on the null hypersurface $u=0$.

Consider a particle with 4-velocity $\bar{u}^\mu$ before it meets the shockwave.
According to geodesic equations the velocity changes due to interaction with the shock.
In the linear in $H$ approximation
\begin{equation}\label{1.2}
	{d u^\mu \over du}=- l^\mu (H_{,\nu}\bar{u}^\nu)-\frac 12 H^{,\mu}
	+O(H^2)\,,
\end{equation}
where we took into account that $(\bar{u}\cdot l)=-1$. It follows from \eqref{1.3}, \eqref{1.2} that the velocity acquires
two parts $u^\mu\simeq \bar{u}^\mu+\chi(u) u^\mu_1+\bar{\chi}(u)u^\mu_2$, where
\begin{equation}\label{1.4}
	u^\mu_1=-\frac 12 l^\mu f\,,\quad u^\mu_2=-\frac 12  l^\mu (f_{,\nu}\bar{u}^\nu)-\frac 12  f^{,\mu}\,.
\end{equation}
The term $\chi(u) u^\mu_1$ in the velocity implies that the advanced time on the particle trajectory changes as
$\delta v(u)=-\bar{\chi}(u)f$.  Terms proportional to $\chi(u)$ disappear after a time  interval exceeding $\delta$.

We suppose that the profile function changes slowly during the action of the shock, $ |f_{,\mu}|\delta \ll |f|$.  Then residual transformations of the particle coordinates and the velocity just after the shock can be written as a pure coordinate transformation
\begin{equation}\label{1.5}
	\delta x^\mu=-\zeta^\mu\,,\quad u^\mu\simeq \bar{u}^\mu+{\cal L}_\zeta \bar{u}^\mu\,,\quad u\simeq \delta\,,
\end{equation}
generated by a specific vector field:
\begin{equation}\label{1.14}
	\zeta^\mu(x)\simeq f\delta^\mu_v+\frac u2 \eta^{\mu a}f_{,a}\,.
\end{equation}
Here ${\cal L}_\zeta$ is the Lie derivative, vector $\zeta^\mu$ has been introduced in \cite{Blau:2015nee}. To derive \eqref{1.14}
we took into account that, in accord with the chosen normalization, $\bar{\chi}(u)=1$ at $u>\delta$.
Coordinate transformations at $u=0$ are well-known Penrose supertranslations  \cite{Penrose:1972xrn}. Also they make an infinite
dimensional group of isometries of the surface $u=0$, known as
Carroll transformations \cite{Duval:2014lpa,Duval:2014uoa,Ciambelli:2023tzb,Ciambelli:2023xqk}.

Thus the gravitational memory effect after particles motion through the shockwave has a pure geometrical interpretation  determined by the profile of the wave $f$. The important fact is that particles do not `remember' the form of the gravitational signal $\chi(u)$.

\section{Gravitational memory of EM fields}  
\label{sec:3}

Our primary interest is to determine perturbations of classical EM fields caused by shockwaves by using the gravitational
memory effect.
Consider Maxwell equations on shockwave spacetime \eqref{1.1}  and in Minkowsky spacetime
\begin{equation}\label{1.15}
	\partial_\mu F^{\mu\nu}=j^\nu\,,\quad\partial_\mu \bar{F}^{\mu\nu}=\bar{j}^\nu \,,
\end{equation}
respectively.  As a result of the specific structure of \eqref{1.1}, $\det g_{\mu\nu}=-1/4$ and both equations look similar,
but indexes of $F^{\mu\nu}$ are risen with the help of the inverse metric tensor $g^{\mu\nu}=\eta^{\mu\nu}+H l^\mu l^\nu$.
Also conservations laws for the currents are: $\partial_\mu j^\mu=\partial_\mu \bar{j}^\mu=0$. We discuss concrete form of the currents latter.

We consider $\bar{F}^{\mu\nu}$ as the Maxwell strength of EM of a source before the shock and define its perturbation as the difference  $\hat{F}_{\mu\nu}\equiv F_{\mu\nu}-\bar{F}_{\mu\nu}$. One can show that \eqref{1.15} lead to equations
\begin{equation}\label{1.16}
	\partial_\mu \hat{F}^{\mu\nu}=J^\nu+\chi(u) J^\nu_g(F,f)\,,
\end{equation}
where $J^\nu=j^\nu-\bar{j}^\nu$ is a variation of the current during the shock, and
\begin{equation}\label{1.17}
	J^\nu_g(F,f)=\partial_\mu\left[f l^\alpha (l^\nu \eta^{\mu\beta}-l^\mu \eta^{\nu\beta}) F_{\alpha\beta})\right]\,.
\end{equation}
It is clear that 
\begin{equation}
	\partial_\mu J^\mu=\partial_\mu (\chi(u) J^\mu_g)=\partial_\mu J^\mu_g=0\,,\quad l_\mu J^\mu_g=0\,.
\end{equation}
In the first approximation one can put
$J^\nu_g(F,f)\simeq J^\nu_g(\bar{F},f)$. After some algebra one can write
\begin{equation}\label{1.17a}
	J^\nu_g(\bar{F},f)=- 2f l_\mu {\cal L}_\zeta \bar{F}^{\mu\nu}-fl^\nu l_\alpha \partial_\lambda \bar{F}^{\alpha\lambda}\,.
\end{equation}
One can interpret $\chi(u) J^\nu_g(\bar{F},f)$ as an effective current, tangent to constant $u$ null hypersurfaces in a narrow region
$0 < u < \delta$ of GSW pulse.
This current is generated by rapidly changing gravitational field on a smooth EM background $\bar{F}_{\mu\nu}$. As we see in a moment, it is
$J^\nu_g(\bar{F},f)$ which determines properties of $\hat{F}^{\mu\nu}$ in a form of EM radiation. The mechanism which yields $\hat{F}^{\mu\nu}$ is analogous to the well-known generation of transition radiation when a charged particle crosses the boundary of two physical media which differ by permittivity or permeability, for example.
Therefore we call $\hat{F}^{\mu\nu}$ transition radiation generated by gravitational shockwaves. Some analogies of the both types of the transition radiation are listed in table \ref{Tab1}.

\begin{table}
	\centering
	\resizebox{\columnwidth}{!}
	{
		\begin{tabular}{|>{\centering}p{.27\linewidth} ||>{\centering}p{.36\linewidth} |>{\centering}p{.38\linewidth}|}
			\hline
			\textbf{EM transition radiation} & \textbf{In media} &  \textbf{On GSW}
			\tabularnewline
			\hline \hline
			Region of origin  &   Space-like boundary between two media with
			different permittivity, permeability  &  Narrow region between null hypersurfaces which are
			boundaries of GSW pulse
			\tabularnewline
			\hline
			Mechanism of \linebreak origin &  Induced electric currents on the boundary &  Effective current along the GSW pulse generated by rapidly changing gravity field
			\tabularnewline
			\hline
			Properties  &  Propagates in both directions from the boundary & Forms EM pulse behind GSW pulse
			\tabularnewline
			\hline
		\end{tabular}
	}
	\caption{\small  Analogies between EM transition radiation in media and on GSW.}
	\label{Tab1}
\end{table}

Our aim is to show that the transition radiation
on shockwaves is determined by the gravitational memory effect and is expressed in a form of coordinate transformation \eqref{1.14}.
We suppose that the transition radiation and variation of the currents in the interval $0\leq u \leq \delta$ have the form:
\begin{align}
	\hat{F}^{\mu\nu}&\simeq \chi(u) p^{\mu\nu}+\bar{\chi}(u) q^{\mu\nu}\,, \label{1.18}
	\\
	J^{\mu}&\simeq \chi(u) i_1^{\mu}+\bar{\chi}(u) i_2^{\mu}\,, \label{1.19}
\end{align}
similar to variations of 4-velocities of particles. Here $p^{\mu\nu}$, $q^{\mu\nu}$, $i_k^{\mu}$ are some quantities which change slowly on the above interval.  It is clear that the gravitational memory effect is determined by $q^{\mu\nu}$ and $i_2^{\mu}$.

By requiring that terms at $\partial_u\chi(u)$, $\chi(u)$, $\bar{\chi}(u)$ in the Bianchi identities and in the conservation law $\partial_\mu J^\mu=0$  vanish one gets the restrictions:
\begin{gather}
	p_{(\mu\nu}l_{\rho)}=0\,,\quad p_{(\mu\nu,\rho)}=q_{(\mu\nu}l_{\rho)}\,, \label{1.20}
	\\
	l_\mu i_1^{\mu}=0\,,\quad\partial_\mu i_1^{\mu}=l_\nu i_2^\nu\,,\quad\partial_\mu i_2^{\mu}=0\,, \label{1.21}
\end{gather}
where symbol $(\mu\nu\ldots\rho)$ denotes complete symmetrization of the indices. Indices are risen and lowered with the flat metric.
Substitution of \eqref{1.18}, \eqref{1.19} to Maxwell equations \eqref{1.16} yields:
\begin{align}
	l_\mu p^{\mu\nu}&=0\,, \label{1.22}
	\\
	\partial_\mu p^{\mu\nu}-l_\mu q^{\mu\nu}&= i_1^{\nu}+J_g^\nu\,, \label{1.23}
	\\
	\partial_\mu q^{\mu\nu}&=i_2^\nu\,. \label{1.24}
\end{align}
From \eqref{1.20}, \eqref{1.22} we find
\begin{gather}
	p_{\mu\nu}=\sigma_i(l_\mu e^i_\nu-l_\nu e^i_\mu)\,, \label{1.25}
	\\
	q_{\mu\nu}=e^i_\mu \sigma_{i,\nu}- e^i_\nu\sigma_{i,\mu}+\rho_i(l_\mu e^i_\nu-l_\nu e^i_\mu)\,, \label{1.26}
\end{gather}
where $\sigma_i$ and $\rho_i$ are some functions. Equations  \eqref{1.25}, \eqref{1.26}  imply that
\begin{equation}\label{1.27}
	\partial_\mu p^{\mu\nu}+l_\mu q^{\mu\nu}=-\sigma_{i,i}l^\nu\,,
\end{equation}
which together with \eqref{1.17a}, \eqref{1.25} yields
\begin{gather}
	l_\mu q^{\mu\nu}=l_\mu {\cal L}_\zeta \bar{F}^{\mu\nu} -{\cal J}^\nu\,, \label{1.28}
	\\
	{\cal J}^\nu=-\frac 12 (i_1^\nu+fl^\nu l_\alpha {\bar j}^\alpha+\sigma_{i,i}l^\nu)\,. \label{1.29}
\end{gather}
Also it follows from \eqref{1.24}, \eqref{1.26} that $\partial_v(\sigma_{i,i})=i_2^v$.  Thus the quantity ${\cal J}^\nu$ is solely determined by the current across the shockwave front. The presence of ${\cal J}^\nu$ indicates possible contributions to transition radiation when the source immediately contacts the shockwave.

To summarize the above analysis we conclude, on quite general grounds, that normal components of transition radiation strength
change  as
\begin{equation}\label{1.30}
	l_\mu \hat{F}^{\mu\nu}\simeq \bar{\chi}(u) \left(l_\mu {\cal L}_\zeta \bar{F}^{\mu\nu}-{\cal J}^\nu\right)\,.
\end{equation}
Outside domains where sources contact the shockwave
the change is generated by the Lie derivative along the supertranslation vector \eqref{1.14}. Equation \eqref{1.30} will be the starting point for the subsequent analysis.

\section{Formulation of the problem}  
\label{sec:4}

We are interested in perturbation $\hat{F}_{\mu\nu}\equiv F_{\mu\nu}-\bar{F}_{\mu\nu}$
in the region $u>\delta$, where both $F_{\mu\nu}$ and $\bar{F}_{\mu\nu}$  are solutions to Maxwell equations in Minkowsky space time.
We require that in this region $j^\mu={\bar j}^\mu$. Therefore, $\hat{F}_{\mu\nu}$ obeys homogeneous Maxwell equations.
Our results allow to formulate the following characteristic Cauchy problem:
\begin{gather}
	\partial^\mu \hat{F}_{\mu\nu}= 0\,,\quad u>\delta\,, \label{1.31a}
	\\
	l^\mu \hat{F}_{\mu\nu}=  l^\mu {\cal L}_\zeta \bar{F}_{\mu\nu}\,,\quad u=\delta\,, \label{1.31b}
\end{gather}
which determines evolution of the transition radiation just behind the shockwave front. In what follows we suppose that the width of the signal is small compared to other  scales of the system and put $\delta=0$. This corresponds to idealized GSW with the world trajectory of the front $u=0$.The problem \eqref{1.31a}, \eqref{1.31b} is called characteristic since the Cauchy data are set on the null hypersurface. We do not consider in this work effects related to possible presence of contact
terms ${\cal J}^\nu$ in the Cauchy data, see our comments below. Our definition implies that $i_2=0$ in \eqref{1.19}.

The Cauchy data fix components $\hat{F}_{vi}$ and $\hat{F}_{vu}$. In fact,
$\hat{F}_{uv}$
is not independent since it follows from the Maxwell equations that
\begin{equation}\label{1.32}
	\hat{F}_{uv}=-\frac12\int_{-\infty}^v \partial_i \hat{F}_{vi}~ dv'\,.
\end{equation}
Also it follows from the Bianchi identities that
\begin{equation}\label{1.33}
		\hat{F}_{ui}=\int_{-\infty}^v \left(\partial_u \hat{F}_{vi}+\partial_i \hat{F}_{uv}\right)dv'\,,
		\quad
		\hat{F}_{ij}=\int_{-\infty}^v \left(\partial_i \hat{F}_{vj}-\partial_j \hat{F}_{vi} \right)dv'\,.
\end{equation}
It is assumed that $\bar{F}_{\mu\nu}$ vanish as $v\to-\infty$ and so does $\hat{F}_{\mu\nu}$.
Since the perturbations satisfy homogeneous Maxwell equations, we come to the characteristic problem for the two components
\begin{gather}
	\Box \hat{F}_{vi}=0\,,\quad u>0\,, \label{1.35}
	\\
	\hat{F}_{vi}={\cal L}_\zeta \bar{F}_{vi}=\partial_v( f \bar{F}_{vi})\,,\quad u=0\,.\label{1.36}
\end{gather}
Given its solution the rest components are determined by \eqref{1.32}, \eqref{1.33}.

It should be emphasized that if there are constraints in the model, the data in the Cauchy problem cannot be arbitrary, but must also satisfy these constraints. In the case of electrodynamics, this is fulfilled, which is easy to verify by comparing \eqref{1.31b} with \eqref{1.32}, \eqref{1.33}. Taking into account the constraints when setting the Cauchy problem is a substantial difference between the gauge model and the scalar model considered in \cite{Fursaev:2024czx}.

We solve problem \eqref{1.35}, \eqref{1.36} for transition radiation when $\bar{F}_{\mu\nu}$ is EM field of compact source (a magnetar). It is convenient to choose a frame in which the magnetar is at rest at the origin of coordinates after crossing the shockwave front. This means that in the chosen frame, before crossing the front, the magnetar was moving with some velocity. As was shown in \cite{Fursaev:2024czx} , when calculating the
data on $u=0$  in the linear order, one can equivalently consider the field of the source located at the origin before crossing the front, but redefine the shockwave profile function 
\begin{equation}
	f(x)\to f(x)-f(0)-x^a\partial_a f(0)\,.
\end{equation}
We will imply that this redefinition in \eqref{1.36} is done.

To find the leading effect in the transition radiation of the magnetar on GSW we approximate the field of the magnetar
by the field of a magnetic dipole at the center of coordinates.  The corresponding current is
\begin{equation}
	\bar{j}_\mu(x)=\delta_\mu^i\varepsilon_{ijk} M_j \, \partial_k \delta^{(3)}(\vec{x})\,,
\end{equation}
where $M_i$ is a magnetic moment (here Latin indices correspond to spatial coordinates $x,y_1,y_2$).
Such a current creates a magnetic field with vector-potential
\begin{equation}
	\bar{A}_\mu(x)=\delta_\mu^i\varepsilon_{ijk} M_j \partial_k \phi\,,\quad
	\phi= - 1/(4\pi r)\,,\quad r^2=x^2+y_i^2\,,
\end{equation}
see \cite{Landau2}. The Cauchy data \eqref{1.36} for the transition radiation
are determined by components 
\begin{equation}
	\bar{F}_{vi}(x) = (\partial_v \mathcal{D}_i^b -\partial_i \mathcal{D}_v^b)\partial_b\phi(x)\,,
\end{equation}
where $i$  corresponds to $y_1=y,\,y_2=z$ and for brevity the following notations are used:
\begin{equation} \label{1.40}
		2\mathcal{D}_v^a = M_y \delta_z^a-M_z\delta_y^a\,,
		\quad
		\mathcal{D}_y^a = -M_x \delta_z^a+2M_z \delta_v^a\,,
		\quad
		\mathcal{D}_z^a =M_x \delta_y^a-2M_y \delta_v^a\,.
\end{equation}

It is convenient to use coordinates $(U,r,\Omega)$, where $U=t-r$ and  $\Omega$ is a solid angle which parametrizes direction
of the unit vector $\vec{e}(\Omega) = (x/r, y^i/r)$. The hypersurface $U=0$, $t>0$ is the future light cone of the event when the shockwave hits the magnetar. All time-like trajectories end up inside the cone, $U>0$.
By following a method of \cite{Fursaev:2024czx} one can show that distant observers measure the following large $r$ asymptotic
at $U>0$
\begin{equation}\label{3.2}
	\hat{F}_{\mu\nu}(U,r,\Omega) \simeq \frac{1}{r} a_{\mu\nu}(U,\Omega) + O(r^{-2})\,,
\end{equation}
where $a_{\mu\nu}$ is an amplitude of outgoing  flux of the transition radiation. In principle, $a_{\mu\nu}$ can be found in an analytic
form, which is lengthy and will be presented in a separate publication. There may be additional contributions to the right hand side of \eqref{3.2} from contact terms in the Cauchy data from ${\cal J}^\nu$. These terms appear when the GSW hits the magnetar.

\section{Transition radiation on GSW produced by ultrarelativistic objects and Fast Radio Bursts}  
\label{sec:5}

As a model of a compact ultrarelativistic object we take a massless point-like particle
which moves along the $x$-axis and has the profile function
\begin{equation}
	f(x) =8G E_p \log \sqrt{(y+a)^2+z^2}\,,
\end{equation}
where $E_p$ is the energy of the particle and $a$ is the impact parameter between the particle and the magnetar.

\begin{figure}[t]
	\includegraphics[width=\textwidth]{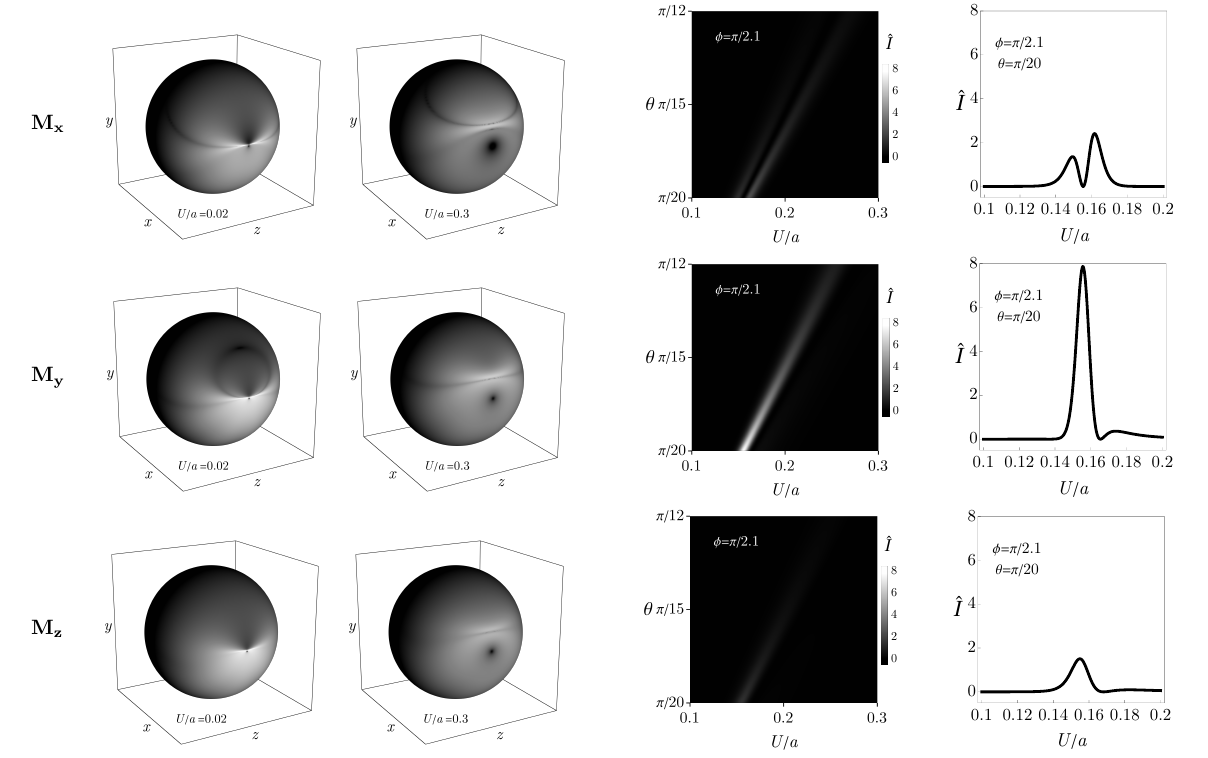}
	\caption{\small  The angular distribution of the transition radiation flux $\hat I(U,\Omega)$ (in dimensionless units) which is generated when the shockwave perturbes the EM field of the magnetar. The source of the shockwave is a massless particle which moves along the $x$-axis. The impact parameter between the particle and the magnetar is $a$.  Orientations of the magnetar's magnetic moment are chosen: toward the particle velocity, $M_x=1$, orthogonal to its velocity in the $y$ and $z$ directions ($M_y=1$, $M_z=1$, respectively). The distributions (two columns on the left) are shown for two moments of time $U$ on a logarithmic scale, the time evolutions (two columns on the right) are shown for fixed spherical angles $\theta,\phi$    for each magnetic moment after the shockwave passed the magnetar.  The intensity of the flux grows in the
		direction of the massless particle velocity, $\theta\to 0$.  Angle $\phi$ changes clockwise in the $y$-$z$ plane. } \label{f2}
\end{figure}

Solution to Cauchy problem \eqref{1.35}, \eqref{1.36} for this profile allows us to find asymptotic \eqref{3.2}
and the angular distribution of the radiation flux 
\begin{equation}
	I(U,\Omega) = r^2 T_U^r (U,r,\Omega)\,,
\end{equation}
where $T_U^r $ are components of the stress-energy tensor of the transition radiation in coordinates $U,r,\Omega$.
The results are  presented on the figure \ref{f2}.

One can point out specific peaks of the intensity, one or two, which appear practically at any observation angle.  The form of peaks depend on orientation of the magnetar's magnetic moment with respect to direction of motion of the GSW. The  duration of the peaks is determined by the impact parameter $a$, it as about $0,01~a$. The height of the peak is proportional to $(GE_pM /a^3)^2$. Figure \ref{f2} shows the dimensionless intensity
$\hat I(U,\Omega)=(a^3/(GE_pM))^2 I(U,\Omega)$. These short pulses of the transition radiation can be the novel engine mechanism for the fast radio bursts as is shown on figure \ref{f1}.

In addition to the impulsive nature of the transition radiation our suggestion is consistent with other requirements.
A model of the synchrotron radiation generated by a pulse of magnetic field coming from a magnetar has been formulated by  Lyubarsky \cite{Lyubarsky:2014jta}. According to this model the amplitude of the magnetic field of the pulse should have the form:
\begin{equation}\label{3.3}
	B_{\mathrm{pulse}}=b B_{*}\frac{R_{*}}{r}\,,
\end{equation}
where $B_{*}$ is the magnetic field on the surface of magnetar, $R_{*}\sim 10^6$ cm is its radius, and $r>10^{16}$ cm is the distance from the magnetar to the region of radio burst formation. The dimensionless parameter $b$ is determined by the primary outburst mechanism. The relation between the surface magnetic field and magnetar's magnetic moments is  $B_{*}\sim M/R_{*}^3$, up to dimensionless factors.

We suppose that the impact parameter between the magnetar and ultrarelativistic object is $a\sim 10~R_{*}\ldots 10^3~R_{*}$. Thus,
the field of transition radiation pulse in the area of radio burst formation satisfies the condition $r\gg a$ and is given by \eqref{3.2}.
This is consistent with \eqref{3.3} for parameter
\begin{equation}\label{3.4}
	b\sim\hat I^{1/2}\frac{GE_p R^2_*}{a^3}\,,
\end{equation}
which should not be too low so that the energy of the primary pulse is sufficient to form the observed FRB. At angles close enough to the direction of motion of the ultrarelativistic object, $\theta\sim \pi/20$, the factor $\hat I(U,\Omega)$ is non-vanishing in the range $U/a\sim 10^{-2}$. For $a\sim 10~R_{*}\ldots 10^3\, R_{*}$ the pulse duration is:
\begin{equation}\label{3.5}
	\delta t\sim 10^{-2}\ldots 1\,\mathrm{ms}\,,
\end{equation}
in accord with the requirements for FRB. The ratio $GE_p/a$ is not small, for example, if the compact object has a mass comparable to that of a neutron star and moves with a Lorentz factor $10$ at a distance $a\sim 10~R_{*}$. In this case $b\sim 10^{-2}$, and the `bright spot' has angle size  $\pi/10$.

\begin{figure}[t]
	\includegraphics[width=\textwidth]{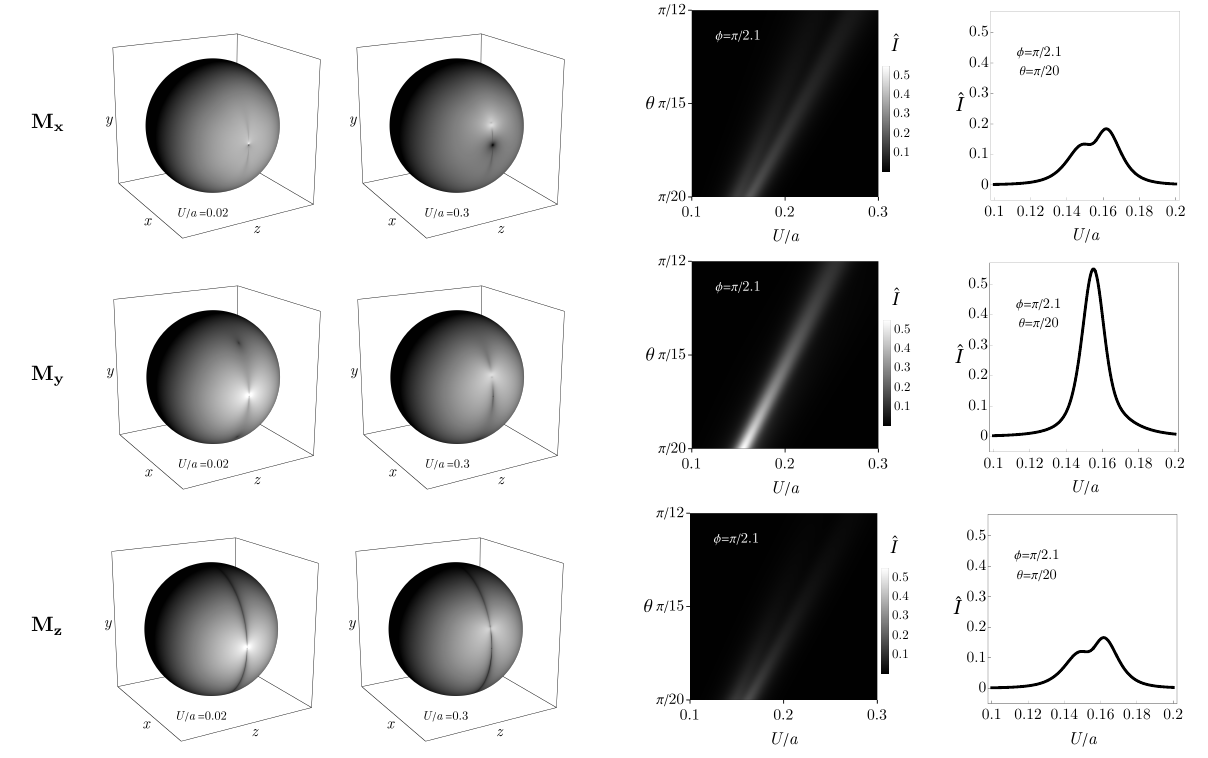}
	\caption{\small shows EM radiation generated by a null cosmic string passing a magnetar. The notations are identical to those
		used for figure \ref{f2}. Although the angular distributions are different the pulses of the transition radiation for GSW from the strings and from compact objects have similar forms. } \label{f3}
\end{figure}

It is also interesting to consider ultrarelativistic cosmic strings as alternative type of objects which produce GSW.
To simplify things one can take a null infinitely thin cosmic string which is directed along the $z$ axis and
moves along the $x$-axis. The corresponding profile function in metric \eqref{1.1} now is 
\begin{equation}
	f(x) =8G\pi E_s |y+a|\,,
\end{equation}
where $E_s$ is the energy of the string per unit length and $a$ is the impact parameter between the string and the magnetar.
For infinitely thin null cosmic string, when $\chi(u)=\delta(u)$ in \eqref{1.1}, the spacetime geometry is locally flat everywhere except singular points on the string worldsheet. Although the geometry is not formally a GSW spacetime computations are the same.
The fact that null cosmic strings passing pulsars  generate EM pulses has been first demonstrated in  \cite{Fursaev:2023lxq} by a similar method based on holonomies of the string spacetime. Here we present computations by using Eqs. \eqref{1.35}, \eqref{1.36}, see figure \ref{f3}. More useful physical estimates can be found in \cite{Fursaev:2022ayo,Fursaev:2023lxq}.
The important conclusion from figure \ref{f3} is that ultrarelativistic cosmic strings  passing magnetars also generate pulses of transition radiation with duration \eqref{3.5} and may ensure the FRB engine mechanism.

It is possible to roughly estimate the energy budget of the transition radiation. For a GSW produced by a
compact object, if the magnetic energy of a magnetar is about $10^{48}$ erg, the power of the transition radiation in the `bright spot' is
\begin{equation}\label{3.6}
	I\sim \left(\frac{GE_p}{a}\right)^2 10^{48}\,\mathrm{erg \cdot s^{-1}}\,,
\end{equation}
for $a\sim 10^7$cm. Take a low-energy FRB power of about $10^{38}\,\mathrm{erg \cdot s^{-1}}$ \cite{Bochenek:2020zxn} and assume 100\% conversion of the transition radiation energy to FRB. Then Eq.~\eqref{3.6} implies that $GE_p/a\gtrsim 10^{-5}$.
For the case of null cosmic strings a similar estimate yields the constraint  $GE_s\gtrsim 10^{-5}$.

\section{Final remarks}  
\label{sec:6}

The aim of the present work was to suggest a novel FRB engine mechanism based on the action of gravitational shockwaves on strong magnetic fields. We consider our estimates as tentative, whose purpose was to reveal the basic features of the suggested mechanism. To proceed with quantitative analysis a detailed picture of the EM pulse interaction with the matter surrounding the magnetar is required.

Also we have not discussed other possible physical properties of
the transition radiation on GSW. It has long been established
\cite{Lichnerowicz1,Lichnerowicz2,Lichnerowicz3,Modugno:1976uf,Modugno:1979rn}
that a plane gravitational shockwave generates an accompanying
electromagnetic shockwave at $u=0$. However, in addition to this
shock, as well as the outgoing radiation flux described in our
work, other physical effects can potentially be observed. The
analysis of scalar models \cite{Fursaev:2024czx} suggests that the
transition radiation may include additional EM shock along the
light cone $U=0$. In particular, if one extends (\ref{3.2}) to
domain $U<0$ there appear contributions to the right hand side of
(\ref{3.2}) from contact terms ${\cal J}^\nu$ to the Cauchy data.
These terms are generated by interaction of the GSW with the
magnetar itself and come out as contributions which are
non-analytical on  $U=0$ (if the magnetar is modelled by a
point-like magnetic dipole). We leave analysis of these effects
for future work.

The transition radiation is not restricted to electromagnetic fields. A null cosmic string which moves near a compact massive object generates gravitational waves, see \cite{Fursaev:2023oep}. It would be interesting to see if there is a gravitational type transition radiation
when gravitational shockwaves perturb gravitational fields of massive objects.

\acknowledgments

The authors are very grateful to Prof. Ilfan Bikmaev for bringing their attention to the problem of FRB, also
they thank Dr. Irina Pirozhenko for valuable discussions.

\end{document}